\documentclass[12pt]{article}

\usepackage{amsmath,amsfonts,amsthm,amssymb,graphicx,epic,eepic,multicol}

\theoremstyle{definition}
\newtheorem{definition}{Definition}
\theoremstyle{remark}
\newtheorem{remark}[definition]{Remark}
\newtheorem{example}[definition]{Example}

\newtheoremstyle{mytheorem}{0.5cm}{0.2cm}{\slshape}{ }{\bfseries}{.}{ }{}
\theoremstyle{mytheorem}

\newcommand{\E}{\mathbf{E}}

\newcommand{\R}{\mathbb{R}}

\DeclareMathOperator{\one}{{1\hspace*{-0.55ex}I}}

\renewcommand{\P}{\mathbf{P}}

\newcommand{\Q}{\mathbf{Q}}

\newcommand{\sT}{\mathcal{T}}

\newcommand{\fF}{\mathfrak{F}}
\newcommand{\salg}{\fF}

\renewcommand{\phi}{\varphi}
\renewcommand{\kappa}{\varkappa}

\newcommand{\nuQ}{\nu}

\newcommand{\thf}{\frac{1}{2}\,}

\newlength{\querylen}
\usepackage{fancybox}

\numberwithin{equation}{section}
\numberwithin{definition}{section}

\begin{document}
\bibliographystyle{plain}

\title{Semi-static hedging for certain Margrabe type options with barriers}

\author{Michael Schmutz\\
  \small Department of Mathematical Statistics and Actuarial Science,\\
  \small University of Bern, Sidlerstrasse 5, 3012 Bern, Switzerland\\
  \small (e-mail: michael.schmutz@stat.unibe.ch)
}

\maketitle

\begin{abstract}
  It turns out that in the bivariate
  Black-Scholes economy Margrabe type options exhibit symmetry properties leading
  to semi-static hedges of rather general barrier options. Some of the results are
  extended to variants obtained by means of Brownian subordination. In order to increase
  the liquidity of the hedging instruments for certain
  currency options, the duality principle can be applied to set up hedges in a foreign
  market by using only European vanilla options sometimes along with a risk-less bond.
  Since the semi-static hedges in the Black-Scholes economy are exact, closed form
  valuation formulas for certain barrier options can be easily derived.

  \medskip


\end{abstract}

\newpage

\section{Introduction}
\label{sec:introduction}

Following Carr and Lee~\cite{car:lee08}, \emph{semi-static
hedging} is the replication of contracts by trading European-style
claims at no more than two times after inception. Semi-static
hedging strategies have been analysed extensively during more than
the last decade, see
e.g.~\cite{and01,and:and:eli02,bow:car94,car:cho97,car:cho02,car:ell:gup98}
and more recently~\cite{car:lee08}.  These strategies are usually
based on European put-call symmetry results, see
e.g.~\cite{bat91,bat97,bow:car94,car94,car:lee08,faj:mor08b,faj:mor06b,faj:mor08,faj:mor08c,pou06}.
Multivariate extensions of the put-call symmetry and the
semi-static hedges can be found in~\cite{mol:sch08}.

Interestingly, also the well-developed \emph{duality principle} in
option pricing traces some of its roots to the same papers as
put-call symmetry results, see e.g.~\cite{bat91,bat97,car94,grab83}.
The power of duality lies in the possibility to reduce the
complexity of various valuation problems by relating them to easier
problems in other, so-called dual markets. For a presentation of
this principle in a general univariate exponential semimartingale
setting see~\cite{eber:pap:shir08}, for bivariate L\'evy markets
see~\cite{faj:mor06c}, for multivariate semi-martingale extensions
(with various dual markets) see~\cite{eber:pap:shir08b}. Certain
typical ideas which are subsumed today under the title of the
duality principle are already discussed in~\cite{mer73}.

Certainly one of the first important applications of the ideas
behind the duality principle was the derivation of the value of an
option to exchange one risky asset for another (also known as
swap-option) by Margrabe in a bivariate Black-Scholes economy,
see~\cite{mar78}.

Despite of the huge amount of literature about financial symmetries
and applications in the univariate case, there seems to be very
little work about multivariate symmetries. In this paper it is shown
that in the bivariate Black--Scholes economy (and partially in
certain related variants obtained by means of Brownian
subordination) Margrabe type options exhibit symmetry properties,
which can be used for deriving semi-static hedging strategies. In
the Black--Scholes case, where the hedging strategies are exact,
these observations can be used as basis for deriving closed-form
formulas for several path-dependent options in an easy way. By
applying the duality principle not only for valuation but also for
implementing semi-static hedges for certain currency options in a
foreign derivative market, the complexity of the instruments in the
hedge portfolios can be reduced to European vanilla options
sometimes combined with a risk-less bond. This also shows that the
analysis of real existing dual markets and their relations to the
original market seems to be even more important than the analysis of
theoretical dual markets.

The importance of developing robust hedging strategies for
multi-asset path-dependent products is particularly stressed by Carr
and Laurence in~\cite{car:laur09}. More specifically we refer to
Haug and Haug~\cite{hau:hau02}, where it is explained how Margrabe
options with knocking features on the ratio process are implicit in
many mergers and acquisitions deals.

The paper is organised as follows: In Section~\ref{sec:add-sym}
symmetry properties of prices of exchange options in a
Black--Scholes economy are analysed and extended to certain other
payoff functions. In Section~\ref{sec:add-sym-time-change} we
discuss markets obtained by means of Brownian subordination
exhibiting a symmetry before we apply symmetries in
Section~\ref{sec:hedge} to hedge certain knock-in and knock-out
options. Based on the derived hedging strategies we show in
Section~\ref{sec:valuation} how certain bivariate path-dependent
valuation problems (in the bivariate Black--Scholes setting) can be
considerably simplified and it is shown, with the help of two
illustrative examples, how closed form formulas for certain concrete
options can be derived in an easy way.

\section{Symmetry in the Margrabe formula}
  \label{sec:add-sym}

In the following we consider vectors as rows or columns depending
on the situation. The imaginary unity is denoted by $i=\sqrt{-1}$
and for two possibly complex vectors we write $\langle
w,z\rangle=w_1z_1+w_2z_2$, $w,z\in\mathbb{C}^2$. The Euclidean
norm for vectors $x\in\R^2$ will be denoted by $\|\cdot\|$.

Let the process $\xi_t=(\xi_{t1},\xi_{t2})$, $t\geq 0$ (later, if
no confusion can arise, often denoted by $\xi_t$ for short), be a
$2$-dimensional $\P$-Brownian motion defined on a filtered
probability space $(\Omega,\salg,(\salg_t)_{t\geq 0},\P)$ (where
$(\salg_t)_{t\geq 0}$ satisfies the usual conditions) with drift
given by
  \begin{equation}
    \label{eq:drift}
       \mu=-\Big(\frac{\sigma_1^2}{2},\frac{\sigma_2^2}{2}\Big)
  \end{equation}
  and the nonsingular covariance matrix being
  \begin{equation}
    \label{eq:centered-gaussian}
       \Sigma=
       \begin{pmatrix}
       \sigma_1^2 & \varrho\sigma_1\sigma_2 \\
       \varrho\sigma_1\sigma_2 & \sigma_2^2
     \end{pmatrix}\,.
  \end{equation}

Define the price processes
\begin{displaymath}
   S_{ti}=S_{0i}e^{\lambda_i t}e^{\xi_{ti}}\,,
   \ t\geq 0\,,\ \lambda_i=r-r_i\,,\ S_{0i}>0\,,\ i=1,2\,.
\end{displaymath}
In currency trading $r_i$ denotes the risk-free interest rate in
the foreign market $i$, while in the share case it becomes the
dividend yield of the $i$th share. Note that $e^{\xi_{t1}}$ and
$e^{\xi_{t2}}$ are $\P$-martingales, as it is the case in the
risk-neutral setting.

We shortly give the dual market derivation presented
in~\cite{eber:pap:shir08b} for this special case in order to
explain the relationship between the Margrabe formula and Bates'
rule. The expectation with respect to probability measure $\Q$ is
denoted by $\E_{\Q}$ and we omit $\Q$ if the expectation is taken
with respect to $\P$.

Since the process $e^{\xi_{t1}}$ is a $\P$-martingale with $\E
e^{\xi_{t1}}=1$ for every $t$, we can introduce a new measure
$\P^1$ by its restrictions to $\salg_t$, given by
\begin{displaymath}
  \frac{d\P^1|_{\salg_t}}{d\P|_{\salg_t}}=e^{\xi_{t1}}\,.
\end{displaymath}
The $1$-dimensional process $e^{\tilde\xi_t}$, with
$\tilde\xi_t=\xi_{t2}-\xi_{t1}$, is a $\P^1$-geometric Brownian
motion and at the same time a $\P^1$-martingale. The parameters of
the Brownian motion with drift are given by
\begin{equation}
  \label{eq:dual-market-BM-parameters}
   \tilde\mu=-\thf\tilde\sigma^2\,,
   \quad
   \tilde\sigma^2=\sigma_1^2+\sigma_2^2-2\varrho\sigma_1\sigma_2\,,
\end{equation}
see e.g.~\cite{eber:pap:shir08b,mar78} or directly note that the
characteristic function of $\tilde\xi_t$ under $\P^1$ for $u\in\R$
is given by
\begin{displaymath}
  \varphi_{\tilde\xi_t}^{\P^1}(u)=\E_{\P^1} e^{iu\tilde\xi_t}
  =\E e^{i\langle(-u-i,u),(\xi_{t1},\xi_{t2})\rangle}
  =e^{t (-iu\thf\tilde\sigma^2-\thf u^2\tilde\sigma^2)}\,,
\end{displaymath}
where~\cite[Th.~25.17]{sat99} is used for the needed extension of
the characteristic function of $\xi_t$. In this dual market we
denote the price ratio process by
\begin{equation}
  \label{eq:U}
  \tilde S_t=\frac{S_{t2}}{S_{t1}}
  =\tilde S_0e^{\tilde\lambda t}e^{\tilde\xi_t}\,,
\end{equation}
where $\tilde S_0=\frac{S_{02}}{S_{01}}$, $\tilde\lambda
=r_1-r_2$, and $\tilde\xi_t$ is defined above. If the process
$e^{\xi_{t2}}$ is taken as the density process for deriving the
measure $\P^2$ from $\P$, an analogous calculation yields that
$e^{-\tilde\xi_t}$ is a $\P^2$-geometric Brownian motion and at
the same time a $\P^2$-martingale, where the parameters of
$-\tilde\xi_t=\xi_{t1}-\xi_{t2}$ under $\P^2$ are again given
by~(\ref{eq:dual-market-BM-parameters}). This phenomenon already
reflects the self-duality of $e^{\tilde\xi_t}$ resulting in the
well-known Bates' rule. If we write shortly
$F_i=S_{0i}e^{\lambda_i T}$, $i=1,2$, and $\tilde F=\tilde
S_0e^{\tilde\lambda T}$ for a finite maturity time $T>0$, then
\begin{align}
  \label{eq:bates-1}
  e^{-rT}\E(aS_{T1}-bS_{T2})_+&=e^{-rT}\E(aF_1e^{\xi_{T1}}-bF_2e^{\xi_{T2}})_+\nonumber\\
  &=S_{01}e^{-r_1T}\E_{\P^1}(a-b\tilde Fe^{\tilde\xi_T})_+\,,
\end{align}
where $a, b\geq 0$, and $x_+=\max(x,0)$ for a real number $x$.
Similarly
\begin{equation}
  \label{eq:bates-2}
  e^{-rT}\E (aF_1e^{\xi_{T2}}-bF_2e^{\xi_{T1}})_+
  =S_{01}e^{-r_1T}\E_{\P^1}(a e^{\tilde\xi_T}-b\tilde F)_+\,.
\end{equation}
By Bates' rule, see e.g.~\cite{bat97}, or by directly applying the
Black-Scholes put- and call-formula for $a,b>0$, we easily obtain
that the right hand sides of~(\ref{eq:bates-1})
and~(\ref{eq:bates-2}) coincide (what they obviously do if one or
both weights vanish) and so do the left hand sides. Hence, for all
$a$, $b\geq 0$,
\begin{equation}
  \label{eq:margr-symmetry}
   e^{-rT}\E(aF_1e^{\xi_{T1}}-bF_2e^{\xi_{T2}})_+
   =e^{-rT}\E(aF_1e^{\xi_{T2}}-bF_2e^{\xi_{T1}})_+
\end{equation}
holds, or in the dual market
\begin{equation}
  \label{eq:dual-mark-pcs}
  S_{01}e^{-r_1T}\E_{\P^1}(a-b\tilde Fe^{\tilde\xi_T})_+
     =S_{01}e^{-r_1T}\E_{\P^1}(ae^{\tilde\xi_T}-b\tilde F)_+\,.
\end{equation}
One can also directly prove~(\ref{eq:margr-symmetry}) with the
help of the Margrabe formula and then obtain Bates' rule as a
corollary.\footnote{We thank an anonymous referee for this hint.}
Note that~(\ref{eq:dual-mark-pcs}) represents arbitrage-free
prices, since $e^{\tilde\xi_T}$ stems from a $\P^1$-martingale.
Thus, we arrive at a symmetry result~(\ref{eq:margr-symmetry}) in
the bivariate Black-Scholes economy being closely related to
Bates' rule~(\ref{eq:dual-mark-pcs}) (often called (classic)
put-call symmetry) in certain dual markets. Note that the symmetry
properties~(\ref{eq:margr-symmetry}) and~(\ref{eq:dual-mark-pcs})
are not time-dependent, i.e.\ they hold for all $t\in[0,T]$.

Note furthermore that for $a<0$, $b>0$,~(\ref{eq:dual-mark-pcs})
trivially holds. From~\cite[Th.~4.1, Remark~4.2]{mol:sch08} (first
time implicitly given in~\cite{car:lee08}) it is known
that~(\ref{eq:dual-mark-pcs}) implies for any integrable payoff
function
\begin{equation}
  \label{eq:self-sual-gen-payoff}
  S_{01}e^{-r_1T}\E_{\P^1}f(\tilde Fe^{\tilde\xi_T})
  =S_{01}e^{-r_1T}\E_{\P^1}\Big[e^{\tilde\xi_T}
  f\big(\frac{\tilde F}{e^{\tilde\xi_T}}\big)\Big]\,.
\end{equation}

Now consider a positive $1$-homogeneous bivariate payoff function
$g$ with positive arguments. A function $g(x,y)$ is called positive
$1$-homogeneous if $g(cx,cy)=cg(x,y)$ for all $c>0$ and $x,y>0$.
Assume that $g(S_{T1},S_{T2})$ is integrable. In view
of~(\ref{eq:self-sual-gen-payoff}) we write this payoff in the form
\begin{equation}
  \label{eq:functions-with-possible-lift}
  g(S_{T1},S_{T2})=S_{T1}g^*\big(\frac{S_{T2}}{S_{T1}}\big)\,,
\end{equation}
where $g^*$ is a function only depending on the price ratio.

By~(\ref{eq:self-sual-gen-payoff}) we get
\begin{align}
  e^{-rT}\E\, &g(F_1e^{\xi_{T1}},F_2e^{\xi_{T2}})\nonumber\\
  &=e^{-rT}\E\Big[F_1e^{\xi_{T1}}g^*\big(\frac{F_2e^{\xi_{T2}}}{F_1e^{\xi_{T1}}}\big)\Big]
   =S_{01}e^{-r_1T}\E_{\P^1}\big[g^*(\tilde Fe^{\tilde\xi_T})\big]\nonumber\\
  &=S_{01}e^{-r_1T}\E_{\P^1}\Big[e^{\tilde\xi_T} g^*\big(\frac{\tilde F}{e^{\tilde\xi_T}}\big)\Big]
   =e^{-rT}\E\Big[F_1e^{\xi_{T2}}g^*\big(\frac{F_2e^{\xi_{T1}}}{F_1e^{\xi_{T2}}}\big)\Big]\nonumber\\
   \label{eq:sym-1-hom}
  &=e^{-rT}\E\, g(F_1e^{\xi_{T2}},F_2e^{\xi_{T1}})\,.
\end{align}

It is known~\cite{car:cho97,car:cho02,car:lee08,pou06} that the
univariate Black-Scholes setting exhibits an additional symmetry.
For integrable payoff functions this can be written as
\begin{equation}
  \label{eq:dual-market-qsd}
  S_{01}e^{-r_1T}\E_{\P^1}f(\tilde S_0e^{\tilde\lambda T+\tilde\xi_T})
  =S_{01}e^{-r_1T}\E_{\P^1}\Big[f\big(\frac{\tilde S_0}{e^{\tilde\lambda
  T+\tilde\xi_T}}\big)\big(e^{\tilde\lambda T+\tilde\xi_T}\big)^\alpha\Big]
\end{equation}
for $\alpha=1-\frac{2\tilde\lambda}{\tilde\sigma^2}$, where we
assume $\tilde\sigma^2>0$. This property is called quasi-self
duality in~\cite{mol:sch08}.

By changing measure and~(\ref{eq:dual-market-qsd}) we get
\begin{align}
  e^{-rT}\E &g(S_{T1},S_{T2})
   =e^{-rT}\E\big[S_{01}e^{\lambda_1T+\xi_{T1}}
   g^*\big(\frac{S_{02}e^{\lambda_2T+\xi_{T2}}}{S_{01}e^{\lambda_1T+\xi_{T1}}}\big)\big]\nonumber\\
  &=S_{01}e^{-r_1T}\E_{\P^1}\big[g^*(\tilde S_0e^{\tilde\lambda T+\tilde\xi_T})\big]
   =S_{01}e^{-r_1T}\E_{\P^1}\big[g^*\big(\frac{\tilde S_0}{e^{\tilde\lambda T+\tilde\xi_T}}\big)
  \big(e^{\tilde\lambda T+\tilde\xi_T}\big)^\alpha\big]
  \nonumber\\
  &=e^{-rT}\E\Big[\big(\frac{S_{01}S_{T2}}{S_{02}S_{T1}}\big)^{\alpha-1}S_{01}
  e^{\lambda_2T+\xi_{T2}} g^*\big(\frac{S_{02}e^{\lambda_1T+\xi_{T1}}}{S_{01}
  e^{\lambda_2T+\xi_{T2}}}\big)\Big]\nonumber\\
  \label{eq:qsd-in-margrabe-general}
  &=e^{-rT}\E\Big[\big(\frac{S_{01}}{S_{02}}\frac{S_{T2}}{S_{T1}}\big)^\beta
  g\big(\frac{S_{01}}{S_{02}}S_{T2},\frac{S_{02}}{S_{01}}S_{T1}\big)\Big]\,,
\end{align}
where $\beta=\alpha-1=\frac{2(r_2-r_1)}{\tilde\sigma^2}$. Hence,
we arrive at an additional symmetry result for positive
$1$-homogeneous integrable payoff functions in the bivariate
Black--Scholes setting. Applied to exchange
options~(\ref{eq:qsd-in-margrabe-general}) yields that
\begin{equation}
  \label{eq:qsd-in-margrabe}
 e^{-rT}\E(aS_{T1}-bS_{T2})_+=e^{-rT}\E\Big[\Big(\frac{S_{01}}{S_{02}}\frac{S_{T2}}{S_{T1}}\Big)^\beta
  \Big(a\frac{S_{01}}{S_{02}}S_{T2}-b\frac{S_{02}}{S_{01}}S_{T1}\Big)_+\Big]\,,
\end{equation}
for $a,b\geq 0$, i.e.\ we end up with an additional symmetry
result for the Margrabe formula. It is possible to
use~\cite[Th.~4.23]{mol:sch08} to see
that~(\ref{eq:qsd-in-margrabe-general}) is in fact already implied
in~(\ref{eq:qsd-in-margrabe}) since $a,b\geq 0$ are arbitrary.

\begin{remark}
  \label{re:stopp}
  Note that if $\tau$ is a random variable with values in $[0,T]$ and
  independent of $\{(\xi_{t1},\xi_{t2}), t\in [0,T]\}$, then
  $(e^{\xi_{\tau1}},e^{\xi_{\tau2}})$ also satisfies~(\ref{eq:sym-1-hom})
  and~(\ref{eq:qsd-in-margrabe-general}) (for $T$ replaced by $\tau$).
\end{remark}

\section{Building L\'evy driven models with symmetry by Brownian
  subordination}
  \label{sec:add-sym-time-change}

In view of the increasing popularity of L\'evy driven price
processes in asset price modeling, we show how bivariate L\'evy
driven price processes exhibiting a useful symmetry can be obtained
by Brownian subordination.

Consider again the bivariate Brownian motion $\xi_t$, $t\geq 0$,
with drift and covariance matrix~(\ref{eq:drift})
and~(\ref{eq:centered-gaussian}). An increasing L\'evy process
(subordinator) $\sT_t$, $t\geq 0$, with $\sT_0=0$, provides the new
time scale, i.e.\ we introduce a ``new'' process by the formula
$\hat\xi_t=\xi_{\sT_t}$, for details see
e.g.~\cite{bra:shi09,con:tan,sat99}. We assume that the Brownian
motion and the time change are \emph{independent}.

Recall that each bivariate L\'evy process $\zeta_t$, $t\geq 0$, is
characterised by its \emph{generating triplet} $(A,\nu,\gamma)$,
where $A$ is a symmetric non-negative definite $2\times 2$ matrix,
$\gamma\in\R^2$ is a constant vector, $\nu$ is a measure on $\R^2$
satisfying $\nuQ(\{0\})=0$, and
\begin{displaymath}
    \quad\int_{\R^2}\min(\|x\|^2,1)d\nu(x)<\infty\,.
\end{displaymath}

Then $\varphi_{\zeta_t}(u)=\E e^{i\langle
u,\zeta_t\rangle}=e^{t\psi_\zeta(u)}$ with characteristic exponent
\begin{equation}
  \label{eq:levy-k-grund}
  \psi_\zeta(u)=i\langle\gamma,u\rangle-\thf\langle u,Au\rangle
      +\int_{\R^n}(e^{i\langle u,x\rangle}-1
      -i\langle u,x\rangle\one_{\| x\|\leq 1})d\nu(x)\,,
\end{equation}
for $u\in\R^2$.

For subordinators, $A$ vanishes and $\nu$ is supported by the
positive half-line, so that the triplet of $\sT_t$ becomes
$(0,\rho,b)$ and

\begin{displaymath}
  \E e^{u\sT_t}=e^{t(bu+\int_0^\infty(e^{ux}-1)\rho(dx))}\,,\
  \text{ for all } u\leq 0\,,
\end{displaymath}
where the measure $\rho$ satisfies $\int_0^\infty(x\wedge
1)\rho(dx)<\infty$ and $b\geq 0$.

It follows e.g.\ from~\cite[Th.~4.2]{con:tan} that $\hat\xi_t$ is
a L\'evy process with characteristic triplet
\begin{equation}
  \begin{cases}
  \label{eq:biv-triplet}
   A=b\Sigma\,,\\
  \nu(x)=\int_0^{\infty}p_s^{\xi}(x)\rho(ds)\,,\\
  \gamma=b\mu+\int_0^\infty\int_{\|x\|\leq 1}x
  p_s^{\xi}(x)dx\rho(ds)\,,
  \end{cases}
\end{equation}
where $p_s^\xi$ is the probability density of $\xi_s$, i.e.\ of the
bivariate normal law with mean $s\mu$, $\mu$ given
by~(\ref{eq:drift}), and nonsingular covariance matrix $s\Sigma$,
$\Sigma$ from~(\ref{eq:centered-gaussian}). Note that by a slight
abuse of notation we denote by $\nu(x)$, $x\in\R^2$, the density of
the L\'evy measure $\nu$. Furthermore, we will see that the
components of $\gamma$ can be written as
in~(\ref{eq:components-by-martingale}).
%
%

Define the corresponding asset price processes by $\hat
S_{ti}=S_{0i}e^{\lambda_it+\hat\xi_{ti}}$, $t\geq 0$, $i=1,2$.
Furthermore, denote by $\salg^{\sT}$ the $\sigma$-algebra generated
by $\sT_t$, $t\geq 0$. For a finite maturity time $T>0$ and an
integrable positive $1$-homogeneous function $g$ we have
\begin{align}
  e^{-rT}\E g(\hat S_{T1},\hat S_{T2})
  &=e^{-rT}\E\, g(F_1e^{\hat\xi_{T1}},F_2e^{\hat\xi_{T2}})\nonumber\\
  &=e^{-rT}\E(\E(g(F_1e^{\xi_{\sT_{T}1}},F_2e^{\xi_{\sT_{T}2}})|\salg^{\sT}))\nonumber\\
  &=e^{-rT}\E(\E(g(F_1e^{\xi_{\sT_{T}2}},F_2e^{\xi_{\sT_{T}1}})|\salg^{\sT}))\nonumber\\
  \label{eq:time-ch-symmetry}
  &=e^{-rT}\E\, g(F_1e^{\hat\xi_{T2}},F_2e^{\hat\xi_{T1}})\,,
\end{align}
i.e.\ the symmetry property~(\ref{eq:sym-1-hom}) still holds.
Similarly we can obtain $\E e^{\hat\xi_{ti}}=1$, $t>0$, $i=1,2$,
i.e.\ $e^{\hat\xi_{ti}}$ are still martingales, see
e.g.~\cite[Prop.~3.17]{con:tan}.

\begin{remark}
  \label{re:stopp-gen}
  Hence, $\hat\xi_t$ is a bivariate L\'evy process with generating
  triplet given by~(\ref{eq:biv-triplet}), so that for a finite
  maturity time $T>0$, $(e^{\hat\xi_{T1}},e^{\hat\xi_{T2}})$ satisfies
  the symmetry property~(\ref{eq:time-ch-symmetry}), which
  stands for a symmetry in arbitrage free prices. Note that if
  $\tau$ is a random variable with values in $[0,T]$ and
  independent of $\{(\hat\xi_{t1},\hat\xi_{t2}), t\in[0,T]\}$,
  then $(e^{\hat\xi_{\tau 1}},e^{\hat\xi_{\tau 2}})$ still
  satisfies~(\ref{eq:time-ch-symmetry}).
\end{remark}
By the martingale property of $e^{\hat\xi_{ti}}$, $t\geq 0$,
$i=1,2$, we can rewrite the components of $\gamma$ as follows

\begin{equation}
  \label{eq:components-by-martingale}
  \gamma_i
  =-\frac{b\sigma_i^2}{2}-\int_{\R^2}(e^{x_i}-1-x_i\one_{\|x\|\leq
  1})\nu(x)dx\,,\quad i=1,2\,,
\end{equation}
cf.~\cite[Prop.~3.18]{con:tan}, and we have $\int_{\|x\|\geq
1}e^{x_i}\nu(x)dx<\infty$, for $i=1,2$,
see~\cite[Th.~25.17]{sat99}.
Equation~(\ref{eq:components-by-martingale}) can be directly
checked by
\begin{multline*}
  \int_0^\infty\int_{\|x\|\leq 1} x_i p_s^{\xi}(x)dx\rho(ds)+\int_0^\infty
  \int_{\R^2}(e^{x_i}-1-x_i\one_{\|x\|\leq 1})p_s^{\xi}(x)dx\rho(ds)\\
  =\int_0^\infty\int_\R (e^{x_i}-1)\int_{\R}p_s^{\xi}(x)dx\rho(ds)
  =0\,.
\end{multline*}

By using~\cite{sat90} or~\cite[Ex.~7.3]{sat00} we see that the
generating triplet of $\hat\xi_t$, $t\in[0,T]$, under the
dual-market measure $\hat\P^1$ (defined by the density process
$e^{\hat\xi_{t1}}$) is given by $(b\Sigma\,,\bar\nu\,,\bar\gamma)$,
where $\bar\nu$ has the density $\bar\nu(x)=e^{x_1}\nu(x)$ and
$\bar\gamma=(\bar\gamma_1,\bar\gamma_2)$ with
\begin{align*}
  \bar\gamma_1&=\frac{b\sigma_1^2}{2}-\int_{\R^2}(e^{x_1}
     -1-x_1e^{x_1}\one_{\|x\|\leq 1})\nu(x)dx\,,\\
  \bar\gamma_2&=b(\varrho\sigma_1\sigma_2-\thf\sigma_2^2)
     -\int_{\R^2}(e^{x_2}-1-x_2e^{x_1}\one_{\|x\|\leq 1})\nu(x)dx\,.
\end{align*}

By~\cite[Prop.~11.10]{sat99} applied for the $1\times 2$ matrix
$U=(-1,1)$, we obtain that $\hat\xi_{t2}-\hat\xi_{t1}$ under
measure $\hat\P^1$ has the triplet
$(b\tilde\sigma^2,\tilde\nu,\tilde\gamma)$ where $\tilde\sigma^2$
is given by~(\ref{eq:dual-market-BM-parameters}),
\begin{displaymath}
  \tilde\nu(y)=\int_0^\infty\int_{\R}e^{x_1} p_s^\xi(x_1,x_1+y)dx_1\rho(ds)\\
              =\int_0^\infty \tilde p_s(y)\rho(ds)
\end{displaymath}
with $\tilde p_s(y)$ being the density of the normal distribution
with mean $-\thf s\tilde\sigma^2$, and variance $s\tilde\sigma^2$
and

\begin{align*}
  \tilde\gamma&=-\frac{b}{2}(\sigma_1^2+\sigma_2^2-2\varrho\sigma_1\sigma_2)-\int_{\R^2}
  (e^{(x_2-x_1)}-1-(x_2-x_1)\one_{|x_2-x_1|\leq
  1})e^{x_1}\nu(x)dx\\
  &=-\frac{b\tilde\sigma^2}{2}-\int_\R(e^y-1-y\one_{|y|\leq 1})\tilde\nu(y)dy\,.
\end{align*}

Since $e^{\hat\xi_{t2}-\hat\xi_{t1}}$ is a $\hat\P^1$ martingale and
$\tilde\nu(y)=e^{-y}\tilde\nu(-y)$, the results
of~\cite{faj:mor06b,faj:mor08c} show that
$e^{\hat\xi_{T2}-\hat\xi_{T1}}$ satisfies~(\ref{eq:dual-mark-pcs})
and~(\ref{eq:self-sual-gen-payoff}) (with $\P^1$ replaced by
$\hat\P^1$) for a fixed maturity time $T>0$. More directly (but
without getting the dynamics) we can use that we already know
from~(\ref{eq:time-ch-symmetry}) that we can exchange
$e^{\hat\xi_{T1}}$ and $e^{\hat\xi_{T2}}$ in the payoff functions of
weighted exchange options. Since $\hat\P^1$ is defined by the
density process $e^{\hat\xi_{t1}}$ the dual market transform for
weighted exchange options at $T>0$ yields that classic put-call
symmetry holds for arbitrary weights, and
by~\cite[Th.~4.1]{mol:sch08} we again arrive
at~(\ref{eq:self-sual-gen-payoff})  (for any integrable payoff
function), where the expectations are taken with respect to
$\hat\P^1$.

\section{Symmetries applied to semi-static hedging}
\label{sec:hedge}

In this section we mainly analyse semi-static hedges of positive
$1$-homogeneous payoff functions with knock-in or knock-out barrier
on the ratio process. Special cases are knock-in and knock-out
Margrabe (exchange) options mentioned in
Section~\ref{sec:introduction}.

To start with consider a positive $1$-homogeneous integrable
payoff function with knock-in features given by the claims
\begin{equation}
  \label{eq:weighted-X-gen}
  X=g(S_{T1},S_{T2})\one_{\big\{\exists t\in[0,T],\;
  c\,\substack{\leq\\ \geq}\,\frac{S_{t2}}{S_{t1}}\big\}}\,,
\end{equation}
where we handle the two cases of up- respectively down-and-in
feature for the ratio process $S_{t2}/S_{t1}$ by $\substack{\leq\\
\geq}$ at the same time, meaning that the inequality for the
up-and-in case is the upper one and the inequality for the
down-and-in case is the lower one. We assume that for the
up-and-in case the spot ratio lies below and for the down-and-in
case above the barrier, i.e. $c\,\substack{>\\<}\,S_{02}/S_{01}$.

Consider the stopping times $\tilde\tau=\inf\{t:c\,\substack{\leq\\
\geq}\,\frac{S_{t2}}{S_{t1}}\}$ and $\tau=\tilde\tau\wedge T$ with
the corresponding stopping $\sigma$-algebras.

First assume that the carrying costs for both assets are the same,
i.e.~$\lambda_1=\lambda_2=r-r_f=\lambda$ and, to begin with,
assume that the asset prices are driven by a Brownian motion with
parameters given by~(\ref{eq:drift})
and~(\ref{eq:centered-gaussian}). Then we can ``lift'' the
corresponding hedging ideas provided in~\cite{car:lee08} in order
to set up the following hedge
\begin{equation}
  \label{eq:hedge-pf-gen}
  G(S_{T1},S_{T2})
  =g(S_{T1},S_{T2})\one_{\big\{cS_{T1}\substack{\leq\\
  \geq} S_{T2}\big\}}
  +g\big(\frac{S_{T2}}{c},cS_{T1}\big)
  \one_{\big\{cS_{T1}\substack{<\\>}S_{T2}\big\}}\,
\end{equation}
for a replication of $X$. The important point here is, that this
hedge for a path-dependent contract does only depend on the asset
prices at maturity, i.e.\ the hedge is of European type.

To confirm the hedge, first note that since $\xi_t$ has stationary
and independent increments, $(\xi_\tau,\xi_T)$ and
$(\xi_\tau,\xi_\tau+\xi'_{T-\tau})$ share the same distribution,
where $\xi'_t$ is an independent copy of the process $\xi_t$.
Hence, $(S_\tau,S_T)$ and $(S_\tau,S_\tau\circ
e^{\xi'_{T-\tau}}e^{\lambda(T-\tau)})$,
$e^{\xi'_{T-\tau}}=(e^{\xi'_{(T-\tau)1}},e^{\xi'_{(T-\tau)2}})$,
also coincide in distribution, where $\circ$ stands for the
componentwise multiplication. Furthermore, note that the second
summand in~(\ref{eq:hedge-pf-gen}) is positive $1$-homogeneous.
Thus, together with~(\ref{eq:sym-1-hom}), Remark~\ref{re:stopp},
and writing shortly $\tau'=T-\tau$, we have on the event
$\{\tilde\tau\leq T\}$,
\begin{align*}
  &e^{-r\tau'}\E[g(S_{T1},S_{T2})|\salg_\tau]
  =e^{-r\tau'}\E[g(S_{\tau 1}e^{\lambda\tau'}e^{\xi'_{\tau'1}},
    S_{\tau 2}e^{\lambda\tau'}e^{\xi'_{\tau'2}})|S_\tau]\\
    &=e^{-r\tau'}\E[g(S_{\tau 1}e^{\lambda\tau'}e^{\xi'_{\tau'1}},
    S_{\tau 2}e^{\lambda\tau'}e^{\xi'_{\tau'2}})
    \one_{cS_{\tau 1}e^{\lambda\tau'}e^{\xi'_{\tau'1}}
    \substack{\leq\\\geq}S_{\tau 2}e^{\lambda\tau'}e^{\xi'_{\tau'2}}}|S_\tau]\\
    &\quad +e^{-r\tau'}\E[g(S_{\tau 1}e^{\lambda\tau'}e^{\xi'_{\tau'1}},
    S_{\tau 2}e^{\lambda\tau'}e^{\xi'_{\tau'2}})
    \one_{cS_{\tau 1}e^{\lambda\tau'}e^{\xi'_{\tau'1}}
    \substack{>\\<}S_{\tau 2}e^{\lambda\tau'}e^{\xi'_{\tau'2}}}|S_\tau]\\
    &=e^{-r\tau'}\E[g(S_{\tau 1}e^{\lambda\tau'}e^{\xi'_{\tau'1}},
    S_{\tau 2}e^{\lambda\tau'}e^{\xi'_{\tau'2}})
    \one_{cS_{\tau 1}e^{\lambda\tau'}e^{\xi'_{\tau'1}}
    \substack{\leq\\\geq}S_{\tau 2}e^{\lambda\tau'}e^{\xi'_{\tau'2}}}|S_\tau]\\
    &\quad +e^{-r\tau'}\E[g(S_{\tau 1}e^{\lambda\tau'}e^{\xi'_{\tau'2}},
    S_{\tau 2}e^{\lambda\tau'}e^{\xi'_{\tau'1}})
    \one_{cS_{\tau 1}e^{\lambda\tau'}e^{\xi'_{\tau'2}}
    \substack{>\\<}S_{\tau 2}e^{\lambda\tau'}e^{\xi'_{\tau'1}}}|S_\tau]\\
    &=e^{-r\tau'}\E[g(S_{\tau 1}e^{\lambda\tau'}e^{\xi'_{\tau'1}},
    S_{\tau 2}e^{\lambda\tau'}e^{\xi'_{\tau'2}})
    \one_{cS_{\tau 1}e^{\lambda\tau'}e^{\xi'_{\tau'1}}
    \substack{\leq\\\geq}S_{\tau 2}e^{\lambda\tau'}e^{\xi'_{\tau'2}}}|S_\tau]\\
    &\quad +e^{-r\tau'}\E[g\big(\frac{S_{\tau 2}}{c}e^{\lambda\tau'}e^{\xi'_{\tau'2}},
    cS_{\tau 1}e^{\lambda\tau'}e^{\xi'_{\tau'1}}\big)
    \one_{S_{\tau 2}e^{\lambda\tau'}e^{\xi'_{\tau'2}}
    \substack{>\\<}cS_{\tau 1}e^{\lambda\tau'}e^{\xi'_{\tau'1}}}|S_\tau]\\
    &=e^{-r\tau'}\E[g(S_{T1},S_{T2})\one_{\{cS_{T1}\substack{\leq\\ \geq}S_{T2}\}}|\salg_\tau]
     +e^{-r\tau'}\E[g\big(\frac{S_{T2}}{c}, cS_{T1}\big)
       \one_{\{cS_{T1}\substack{<\\>}S_{T2}\}}|\salg_\tau]\,,
\end{align*}
where the third equality holds by symmetry and the fourth since
$cS_{\tau 1}=S_{\tau 2}$ on the event $\{\tilde\tau\leq T\}$.
Thus, whenever the ratio process $S_{t2}/S_{t1}$ of the two assets
is equal to $c$ before $T$, we have shown the zero costs of the
exchange from the hedge-claim $G$ to the needed claim $g$. On the
event $\{\tilde\tau >T\}$, the barrier never knocks in and the
claim $G$ expires worthless, as desired.

\begin{example}
  \label{ex:nice-knockers}
  For concrete and sufficiently simple examples, the complexity of
  the hedge claim can be reduced considerably by elementary arguments.
  For example consider the claims
  \begin{align*}
    X_{\rm{ex}}&=(aS_{T1}-bS_{T2})_+\one_{\exists t\in[0,T]\,,\ cS_{t1}\leq S_{t2}}\,,\\
    Y_{\rm{ex}}&=(aS_{T1}-bS_{T2})_+\one_{cS_{t1}>S_{t2},\ \forall t\in[0,T]\,,}
  \end{align*}
  where $cS_{01}>S_{02}$, $0<a\leq bc$ and
  $\tilde\tau=\inf\{t:c\leq\frac{S_{t2}}{S_{t1}}\}$. Then
  by~(\ref{eq:hedge-pf-gen}) the hedge portfolio for $X_{\rm{ex}}$
  is given by
  \begin{displaymath}
    G(S_{T1},S_{T2})=(aS_{T1}-bS_{T2})_+\one_{cS_{T1}\leq S_{T2}}
    +\big(\frac{a}{c}S_{T2}-bcS_{T1}\big)_+
    \one_{cS_{T1}<S_{T2}}\,.
  \end{displaymath}
  Note that since $0<a\leq bc$ we have that if $\one_{cS_{T1}\leq
  S_{T2}}$ is one at $T$, then $(aS_{T1}-bS_{T2})_+$ is worthless, i.e.\
  the first summand is identically zero. Furthermore, and again by
  $0<a\leq bc$, we have that if  $(\frac{a}{c}S_{T2}-bcS_{T1})_+$
  is (strictly) positive $\one_{cS_{T1}<S_{T2}}$ is automatically one
  so that we can forget the indicator function in the second summand.
  Hence, $X_{\rm{ex}}$ can be hedged by taking a long position in the
  European derivative with payoff $(\frac{a}{c}S_{T2}-bcS_{T1})_+$.
  By the knock-in knock-out parity $Y_{\rm{ex}}$ can be hedged by
  taking a short position in this derivative and a long position in
  the European derivative with payoff $(aS_{T1}-bS_{T2})_+$. This
  can also easily be checked directly by calculating the non-discounted
  prices on the event $\{\tilde\tau\leq T\}$
  \begin{align}
    \label{eq:start-hedge-der-1-gen}
    \E[(aS_{T1}-bS_{T2})_+|\salg_\tau]
     &=\E[(aS_{\tau 1}e^{\lambda\tau'}e^{\xi'_{\tau'1}}
      -bS_{\tau 2}e^{\lambda\tau'}e^{\xi'_{\tau'2}})_+|S_\tau]\\
     \label{eq:start-hedge-der-2-gen}
     &=\E[(\frac{a}{c}S_{\tau 2}e^{\lambda\tau'}e^{\xi'_{\tau' 2}}
       -bcS_{\tau 1}e^{\lambda\tau'}e^{\xi'_{\tau'1}})_+|S_\tau]\\
     &=\E[(\frac{a}{c}S_{T2}-bcS_{T1})_+|\salg_\tau]\nonumber
  \end{align}
  and noticing that $(\frac{a}{c}S_{T2}-bcS_{T1})_+$ expires worthless
  on the event $\{\tilde\tau >T\}$ (since then $S_{T2}<cS_{T1}$ and
  $a\leq bc$).

  A serious problem for implementation would be possible non-liquidity
  of the derivatives in the hedge-portfolio. But if the dual markets
  from Section~\ref{sec:add-sym} exist in reality, we can use the
  duality principle not only for valuation purposes but also for
  better implementation of the hedge portfolios. Obvious examples with
  existing dual markets are currency markets. In such a case $S_{t1}$
  represents the price-process of a first currency, the other is given
  by $S_{t2}$. Then the price of the second currency in the first
  foreign market is represented by the ratio process $\tilde S_t$.

  Recall that in order to get equal carrying costs in the domestic
  market we need $r_1=r_2$ (denoted by $r_f$) for a moment and further recall that
  $e^{\xi'_t}$ is an independent copy of $e^{\xi_t}$. The change of measure
  for the time interval $[\tau,T]$ (on the event $\{\tilde\tau\leq T\}$)
  in~(\ref{eq:start-hedge-der-1-gen})
  respectively~(\ref{eq:start-hedge-der-2-gen}) is given by
  $e^{\xi'_{(T-\tau)1}}$, so that
  \begin{align}
    \label{eq:foreign-hedge-long}
    e^{-r(T-\tau)}\E[(aS_{T1}-bS_{T2})_+|\salg_\tau]
    &=S_{\tau 1}e^{-r_f(T-\tau)}\E_{\P^1}[(a-b\tilde S_T)_+|\salg_\tau]\,,\\
     \label{eq:foreign-hedge-short}
    e^{-r(T-\tau)}\E[(\frac{a}{c}S_{T2}-bcS_{T1})_+|\salg_\tau]
    &=S_{\tau 1}e^{-r_f(T-\tau)}\E_{\P^1}[(\frac{a}{c}\tilde S_T-bc)_+|\salg_\tau]\,.
  \end{align}
  Hence, the value at $\tau\in[0,T]$ of the European derivative with
  payoff $(a-b\tilde S_T)_+$ traded in the first foreign market in the first
  currency changed (by $S_{\tau 1}$) into the domestic currency is
  equal to the long position in our hedging portfolio. Furthermore,
  the value at time $\tau$ of the European derivative with payoff
  $(\frac{a}{c}\tilde S_T-bc)_+$ changed into the domestic currency is equal
  to the short position in the hedging portfolio of $Y_{\rm{ex}}$. By positive
  homogeneity $(a-b\tilde S_T)_+=b(\frac{a}{b}-\tilde S_T)_+$ and
  $(\frac{a}{c}\tilde S_T-bc)_+=\frac{a}{c}(\tilde S_T-\frac{bc^2}{a})_+$ so that the
  hedging strategies can be implemented by using only European vanilla
  options. Analogous results can be derived for hedging in the second
  foreign market and for other positive $1$-homogeneous payoff functions (based
  on~(\ref{eq:hedge-pf-gen})).
  \end{example}

Many well-known other examples of positive $1$-homogeneous payoff
functions $g$ can be found in the rich literature about the
duality principle, see e.g.~\cite{faj:mor08}.

For more complicated situations than presented in
Example~\ref{ex:nice-knockers} it is worth to recall that positive
$1$-homogeneous European payoff functions (contained in the hedges)
can be written as a product of the first asset and a function on the
price ratio at maturity. By this observation we can relate the
functions on the price ratio at maturity to the approximation
approach presented in~\cite{car:lee08} (where approximated
portfolios in the domestic market contain exchange options after
multiplying back by the first asset). For details including some
mild regularity assumptions we refer to~\cite{car:lee08}.

Above we have derived the hedging strategies in a bivariate
Black--Scholes setting with equal carrying costs. However, if we
replace the Brownian motion with drift by the L\'evy process
$\hat\xi_t$ from the construction presented in
Section~\ref{sec:add-sym-time-change}, then we get bivariate
L\'evy driven price process, for which we are still in a
risk-neutral setting, we still have the needed symmetry in the
model and since $\hat\xi_t$ has stationary and independent
increments, $(\hat\xi_\tau,\hat\xi_T)$ and
$(\hat\xi_\tau,\hat\xi_\tau+\hat\xi'_{T-\tau})$ share the same
distribution on the event $\{\tilde\tau\leq T\}$, where
$\hat\xi'_t$ is an independent copy of the process $\hat\xi_t$.
However, a glance to~(\ref{eq:biv-triplet}) shows that we can not
be sure anymore that $cS_{\tau 1}=S_{\tau 2}$ on the event
$\{\tilde\tau\leq T\}$. Due to this fact the resulting hedges are
not accurate anymore,  e.g.\ for the claim $X_{\rm{ex}}$ we end up
with a super-replication, for $Y_{\rm{ex}}$ with a more
problematic sub-replication.

Further hedging strategies presented e.g.\
in~\cite{car:ell:gup98,car:lee08} can also be ``lifted'' in the
obvious way leading to hedging strategies for certain bivariate
options with certain other knock-in and knock-out features for the
ratio process. Knock-in and knock-out features related to the
asset price processes (in the bi- or multivariate setting) or more
complicated ``underlying'' payoff-functions than the above used
positive $1$-homogeneous ones for bivariate, and even more for
multivariate options, need other, usually stronger, symmetry
properties of the driving processes, see~\cite{mol:sch08}.

Unlike equity markets, where the assumption of equal carrying
costs is often not totally unrealistic, e.g.\ in non-dividend
cases, this assumption is quite restrictive in currency markets,
since the risk-free interest rates in two different countries are
usually not the same. In order to get exact hedges in a bivariate
\emph{Black-Scholes economy with different carrying costs} we can
use~(\ref{eq:qsd-in-margrabe-general}) instead
of~(\ref{eq:sym-1-hom}) and rely on the above ideas to see that
the hedge portfolio in~(\ref{eq:hedge-pf-gen}) has to be replaced
by
\begin{equation}
  \label{eq:hedge-gener}
  G(S_{T1},S_{T2})
  =g(S_{T1},S_{T2})\one_{\big\{cS_{T1}\substack{\leq\\
  \geq} S_{T2}\big\}}
  +\big(\frac{S_{T2}}{cS_{T1}}\big)^\beta g\big(\frac{S_{T2}}{c},cS_{T1}\big)
  \one_{\big\{cS_{T1}\substack{<\\>}S_{T2}\big\}}\,,
\end{equation}
where $\beta=\frac{2(r_2-r_1)}{\tilde\sigma^2}$.

For instance $X_{\rm{ex}}$ can be hedged with a long position in
\begin{equation}
  \label{eq:hedge-short-diff-cc}
  (\frac{a}{c}S_{T2}-bcS_{T1})_+
  \Big(\frac{S_{T2}}{cS_{T1}}\Big)^{\frac{2(r_2-r_1)}{\tilde\sigma^2}}\,,
\end{equation}
and the hedge for $Y_{\rm{ex}}$ is given by a long position in the
European derivative with payoff $(aS_{T1}-bS_{T2})_+$ (as in the
equal carrying costs case) and a short position in the European
derivative with payoff function~(\ref{eq:hedge-short-diff-cc}). We
refer to~(\ref{eq:qsd-in-margrabe}) and especially
to~(\ref{eq:value-X}) to see directly that $cS_{\tau 1}=S_{\tau
2}$ on the event $\{\tilde\tau\leq T\}$ implies that the short and
the long position of $Y_{\rm{ex}}$ share the same price. To
implement the hedge of $Y_{\rm{ex}}$ in the first foreign market
we should enter again a long-position in the European derivative
with payoff $(a-b\tilde S_T)_+=b(\frac{a}{b}-\tilde S_T)_+$ and
short the European derivative with payoff
\begin{equation}
  \label{eq:ugl-hedge}
  \big(\frac{\tilde S_T}{c}\big)^{\frac{2(r_2-r_1)}{\tilde\sigma^2}}
    \big(\frac{a}{c}\tilde S_T-bc\big)_+
    =\big(a-\frac{bc^2}{\tilde S_T}\big)_+
    \big(\frac{\tilde S_T}{c}\big)^{1-\frac{2(r_1-r_2)}{\tilde\sigma^2}}\,.
\end{equation}

Since the hedging instrument defined by~(\ref{eq:ugl-hedge}) can
be decomposed in vanilla options and a risk-less bond, see
again~\cite{car:lee08,car:mad94}, also in this more complicated
case it is possible to create a semi-static hedge containing only
standard instruments, i.e.\ European vanilla options and a
risk-less bond.

Hence, well-established and analysed univariate semi-static hedging
strategies can be ``lifted'' in order to get semi-static hedging
strategies for (bivariate) derivatives with positive $1$-homogeneous
payoffs with certain knock-in and knock-out features on the ratio
process and sometimes the hedges can be implemented by using
univariate strategies in a dual market. The effectiveness of these
univariate semi-static hedges based on put-call symmetry was
confirmed in several articles either by simulations or empirical
tests, see e.g.~\cite{car:wu06,nal:pou06,tom02}.

\section{Closed form valuation formula}
\label{sec:valuation}

Since in the Black-Scholes economy the asset price processes have
continuous sample paths, the hedges from Section~\ref{sec:hedge}
for unequal carrying costs are exact. Hence, bivariate
path-dependent valuation problems of the
form~(\ref{eq:weighted-X-gen}) can be reduced to European ones of
the form~(\ref{eq:hedge-gener}) for which the duality principle is
applicable. On the basis of this observation, closed form
valuation formulas can sometimes be derived in an easy way.

E.g.\ for the claims $X_{\rm{ex}}$ and $Y_{\rm{ex}}$ from
Example~\ref{ex:nice-knockers} we can calculate the values of the
European derivatives with payoff functions $(aS_{T1}-bS_{T2})_+$ and
$(\frac{a}{c}S_{T2}-bcS_{T1})_+\big(\frac{S_{T2}}{cS_{T1}}\big)^\beta$,
 $\beta=\frac{2(r_2-r_1)}{\tilde\sigma^2}$, and $0< a\leq bc$,
where we use the notation of the last sections. The value of the
first derivative is obtained by the Margrabe formula. By noticing
that
\begin{displaymath}
  \E\Big((\frac{a}{c}S_{T2}-bcS_{T1})_+
    \big(\frac{S_{T2}}{cS_{T1}}\big)^\beta\Big)
    =\E_{\P^1}\Big((\frac{a}{c}F_2e^{\tilde\xi_T}-bcF_1)_+
      \big(\frac{F_2}{cF_1}e^{\tilde\xi_T}\big)^\beta\Big)
\end{displaymath}
and a standard calculation we arrive at
  \begin{multline}
    e^{-rT}\E\Big((\frac{a}{c}S_{T2}-bcS_{T1})_+
      \big(\frac{S_{T2}}{cS_{T1}}\big)^\beta\Big)\\
     \label{eq:val-compl}
     =\frac{a}{c}S_{02}
     \Big(\frac{S_{02}}{cS_{01}}\Big)^{\frac{2(r_2-r_1)}{\tilde\sigma^2}}
     e^{-r_1T}\Phi\big(\tilde d_2\big)
     -bcS_{01}\Big(\frac{S_{02}}{cS_{01}}\Big)^{\frac{2(r_2-r_1)}{\tilde\sigma^2}}
     e^{-r_2T}\Phi\big(\tilde d_4\big)\,,
  \end{multline}
  where
  \begin{displaymath}
    \tilde d_2=\frac{\log(\frac{aS_{02}}{bc^2S_{01}})
      +(r_2-r_1+\thf\tilde\sigma^2) T}{\tilde\sigma\sqrt
      T}\,,\quad \tilde
      d_4= \tilde d_2-\tilde\sigma\sqrt{T}\,,
  \end{displaymath}
  $\Phi(x)=\frac{1}{\sqrt{2\pi}}\int_{-\infty}^xe^{-\thf
  y^2}\,dy$, being the value of $X_{\rm{ex}}$. For the value $V_0$ of the claim $Y_{\rm{ex}}$
  at time $0$ we arrive at the following formula
  \begin{multline}
    V_0=ae^{-r_1T}\Big(S_{01}\Phi(\tilde d_1)
    -\frac{S_{02}}{c}\left(\frac{S_{02}}{cS_{01}}\right)^{\frac{2(r_2-r_1)}{\tilde\sigma^2}}\Phi(\tilde d_2)\Big)\\
    \label{eq:value-X}
    -be^{-r_2T}\Big(S_{02}\Phi(\tilde d_3)
    -S_{01}c\left(\frac{S_{02}}{cS_{01}}\right)^{\frac{2(r_2-r_1)}{\tilde\sigma^2}}\Phi(\tilde d_4)\Big)\,,
  \end{multline}
  where $\tilde d_2$ and $\tilde d_4$ are as above and
  \begin{displaymath}
    \tilde d_1=\frac{\log(\frac{aS_{01}}{bS_{02}})
      +(r_2-r_1+\thf\tilde\sigma^2) T}{\tilde\sigma\sqrt
      T}\,,\quad \tilde
      d_3=\tilde d_1-\tilde\sigma\sqrt{T}\,.
  \end{displaymath}
This formula is consistent with one of the results presented
in~\cite{hau:hau02} derived with the help of the reflection
principle. More valuation formulas of knock-in/out Margrabe
options, derived with the help of the reflection principle, can
also be found in this article.

\section*{Acknowledgements}
\label{sec:acknowledgements}

The author is grateful to Ilya Molchanov for helpful discussions and
corrections and to Sara Fischer and Olivia Jutzi for proofreading,
to Rolf Burgermeister and Markus Liechti for hints from practice,
and especially to two anonymous referees for numerous inspiring and
helpful comments. This work was supported by the Swiss National
Science Foundation Grant Nr. 200021-117606.

\newcommand{\noopsort}[1]{} \newcommand{\printfirst}[2]{#1}
  \newcommand{\singleletter}[1]{#1} \newcommand{\switchargs}[2]{#2#1}

\end{document}